\documentclass[journal]{IEEEtran}

\usepackage{amsmath}					% amsmath is IMPORTANT

\DeclareMathOperator*{\argmin}{arg\,min}
\usepackage{amssymb}					% amssymb is IMPORTANT
\usepackage{amsthm}					% amsthm is IMPORTANT
\usepackage{amsfonts}					% amsfonts is IMPORTANT
\usepackage{mathrsfs}					% mathrsfs is IMPORTANT
\usepackage{mathtools}					% mathtools is IMPORTANT			
\usepackage{mathdots}					% For dot sign
\usepackage{fancyhdr}					% For head franchise 
\usepackage{longtable}					% For long table
\usepackage{xcolor}						% For colorful expression
\usepackage{multirow, makecell}			% For fancy table cell and multi-row tables
\usepackage{booktabs}					% For convenient booktabs
\usepackage{siunitx}					% For miscellaneous math symbols
\usepackage{nomencl}					% For nomenclature
\usepackage[short]{optidef}				% For definition marks				
\usepackage[utf8x]{inputenc}				% For unicode character
\usepackage{kantlipsum}					% For fake Latin text in Kant style
\usepackage[pdftex]{graphicx} 			% For adding figures (IMPORTANT)
\usepackage{cite}						% For citations (IMPORTANT)
\usepackage{pgfplots}					% For modifying plots and generating figures itself
\usepackage{epstopdf}					% For transferring EPS file to PDF file and adding to LaTeX
\usepackage{listings}					% For listing any code in LaTeX
\usepackage{enumerate}					% For making list (IMPORTANT)
\usepackage{tabularx}					% For making fancy tables (IMPORTANT)
\usepackage[normalem]{ulem}				% For highlighting and emphasis
\usepackage{algorithm, algorithmic}		% For presenting algorithms
\usepackage[flushleft]{threeparttable}		% For miscellaneous alignments
\usepackage{changepage}				% For \adjustwidth command for writing pseudocodes

\usepackage{caption}
\usepackage{hyperref}
\usepackage{array}
\usepackage{booktabs} 
\usepackage{subcaption} % 引入 subcaption 宏包

\graphicspath{ {/Users/Desktop/} }

\begin{document}

\captionsetup[figure]{name={Fig.},labelsep=period}

\title{Model Inversion Attack against Federated Unlearning}

\author{Lei~Zhou, ~Youwen~Zhu

\thanks{
}
}

\markboth{}{}

\maketitle

\begin{abstract}
    With the enactment of regulations related to the ``right to be forgotten", federated learning is facing new privacy compliance requirements. Therefore, federated unlearning (FU) is designed to efficiently remove the influence of specific data from FL models, ensuring that user data can be revoked to enhance privacy.
    However, existing FU studies primarily focus on improving unlearning efficiency, with little attention given to the potential privacy risks introduced by FU itself. 
    To bridge this research gap, we propose a novel federated unlearning inversion attack (FUIA) to expose potential privacy leakage in FU. This work represents the first systematic study on the privacy vulnerabilities inherent in FU.  
    FUIA can apply to three major FU scenarios: sample unlearning, client unlearning, and class unlearning, demonstrating broad applicability and threat potential. Specifically, the server, acting as an honest-but-curious attacker, continuously records model parameter changes throughout the unlearning process and analyzes the differences before and after unlearning to infer the gradient information of forgotten data, enabling the reconstruction of its features or labels. 
    FUIA directly undermines the goal of FU to eliminate the influence of specific data, exploiting vulnerabilities in the FU process to reconstruct forgotten data, thereby revealing flaws in privacy protection.  
    Moreover, we explore two potential defense strategies that introduce a trade-off between privacy protection and model performance.  
    Extensive experiments on multiple benchmark datasets and various FU methods demonstrate that FUIA effectively reveals private information of forgotten data.

\end{abstract}


\IEEEpeerreviewmaketitle

\section{Introduction}
\IEEEPARstart{W}{ith} the rapid development of artificial intelligence, the issue of data privacy and security in training machine learning models has garnered increasing attention from both academia and industry \cite{de2021critical}. To address these concerns, various regulations have been enacted worldwide, such as the General Data Protection Regulation (GDPR) \cite{voigt2017eu} in the European Union and the California Consumer Privacy Act (CCPA) \cite{harding2019understanding}  in the United States. These regulations provide robust legal protections for data owners, with the ``right to be forgotten" \cite{regulation2016regulation} drawing significant attention in recent years. This right mandates that when a data owner requests the deletion of their data, the model owner must not only remove the corresponding data from the database but also eliminate its impact on the trained model.
Federated learning (FL) \cite{learning2017collaborative} is designed to enable distributed model training while preserving the privacy of individual clients. However, the introduction of the ``right to be forgotten" imposes new requirements on FL, particularly in efficiently fulfilling data owners' requests for data removal. Recent studies have proposed a novel approach known as federated unlearning (FU) \cite{liu2024survey}, which aims to erase the influence of the target data from the model without the need for retraining the entire model. 

\begin{figure}[t]
    \centering
    \includegraphics[width=0.46\textwidth]{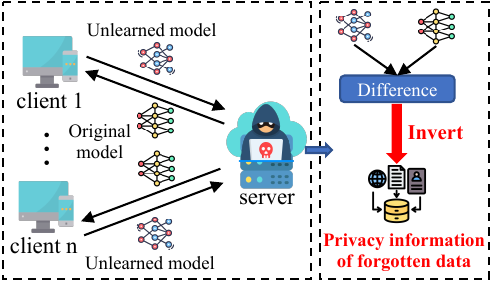} 
    \caption{Overview of FUIA. The server (attacker) inverts by comparing the differences in the model parameters of the stored unlearning process to get the privacy information of the forgotten data.}
    \label{fig_intro}
\end{figure}

Current research on FU primarily focuses on improving the efficiency of the unlearning process while paying little attention to the potential privacy leakage issues inherent in FU \cite{wang2024fedu, shaik2024framu, liu2021federaser, zhang2023fedrecovery, wang2022federated, liu2022right, cao2023fedrecover}.
FedMUA attacks by maliciously exploiting the unlearning mechanism of FU, causing the model to intentionally forget more information, leading to biased prediction results \cite{chen2025fedmua}. However, this attack mainly targets the performance aspect of the model. Overall, current research on FU security is still not comprehensive, particularly lacking in-depth exploration of the privacy vulnerabilities inherent in FU itself.
To fill the research gap, we propose a novel attack method called federated unlearning inversion attack (FUIA). To the best of our knowledge, this is the first study to specifically explore the privacy leakage issues inherent in FU.

During the unlearning process, the server has full access to the parameter differences between the original model and the unlearned model, and it can precisely record the parameter updates uploaded by clients in each training round. However, this extensive access poses significant privacy risks, particularly regarding the leakage of forgotten data.
In FL, prior studies have exploited the server's ability to track model updates and proposed the gradient inversion attack (GIA) \cite{zhu2019deep, liang2023egia}, which reconstructs clients' original training data by analyzing parameter changes. Inspired by this, we introduce a novel attack against FU—federated unlearning inversion attack (FUIA), which aims to infer the features or labels information of forgotten data by leveraging parameter changes during the unlearning process.
FUIA directly compromises the purpose of FU, as forgotten data remains vulnerable to privacy leakage even after unlearning is completed.

In FUIA, the attacker is modeled as an "honest-but-curious" server that follows the prescribed training and unlearning protocols while continuously recording client-uploaded parameter updates and analyzing the parameter differences before and after unlearning. This approach of storing parameter updates has been widely adopted in privacy studies of FL and FU, making it a reasonable assumption rather than an additional constraint.
As shown in Figure  \ref{fig_intro}, the attacker in FUIA exploits the stored differences between the original and unlearned model parameters to estimate the gradient information of forgotten data, thereby reconstructing its features or inferring its labels information.

Based on different unlearning targets, existing research on FU can be classified into three scenarios: sample unlearning, client unlearning, and class unlearning \cite{li2025machine}. Sample unlearning focuses on eliminating the influence of several data samples. Client unlearning requires the model to entirely forget the data from a specific client, making sure that the contributions of the client are no longer reflected in the model parameters. Class unlearning aims to completely remove the influence of a specific class of data from the model. 
To exploit vulnerabilities in these FU methods, we propose FUIA, which targets all three scenarios of unlearning mechanisms. 
(1) In the cases of sample unlearning and client unlearning, FUIA aims to recover the features of the forgotten data, such as reconstructing images. The core principle behind this attack lies in the observation that the parameter differences before and after unlearning inherently encode gradient information related to the forgotten data. By leveraging techniques inspired by GIA, which are well-studied in FL, the attacker can apply optimization-based inversion methods to reconstruct the forgotten data.  
(2) For class unlearning, since the server already possesses knowledge of the global dataset, the goal of FUIA shifts from reconstructing individual data samples to inferring the label of the forgotten class. In general, data from a particular class has a significant impact on model parameters, especially in the output layer. Therefore, the attacker can compare the parameter changes of the output layer before and after unlearning to infer the label of the removed class.
The attack strategy of FUIA demonstrates its broad applicability across different FU scenarios.

Moreover, we propose two general defense strategies. 
First, during the unlearning process in FU, we introduce gradient pruning, where a certain proportion of less important parameter updates from each client is removed before aggregation. The primary goal of this method is to prevent the server from obtaining a complete and precise set of model updates. Second, we leverage differential privacy as an additional defense mechanism. During the unlearning process, we introduce random noise into the model updates submitted by clients, ensuring that the server only receives perturbed parameter updates. 
These approaches effectively both disrupt the ability of the attackers to infer precise gradient information related to the forgotten data, thereby weakening the effectiveness of FUIA. However, they can lead to poor model performance. Therefore, a careful balance must be struck to enhance security while minimizing performance degradation. We hope those approaches will inspire future research in developing more efficient and robust FU mechanisms.

Our contributions can be summarized as follows:  
\begin{enumerate}[1)] 
    \item We are the first to investigate the inherent privacy risks in FU and demonstrate that FUIA can completely undermine the goal of FU by reconstructing the features and labels of forgotten data, leading to severe privacy leakage.  
    \item FUIA exhibits a broad attack scope, effectively targeting three primary FU scenarios: sample unlearning, client unlearning, and class unlearning.
    \item We propose and validate two defense strategies that offer new insights to enhance the privacy protection of FU.
    \item We conduct comprehensive experiments on multiple benchmark datasets across three different FU scenarios, demonstrating the effectiveness of FUIA in exposing the privacy information of forgotten data.
\end{enumerate}

\section{Related work}
\label{relatedwork}
\subsection{Federated Unlearning}
The primary goal of FU is to derive an unlearned model directly from the original model without retraining, thereby significantly improving the efficiency of the unlearning process. Depending on the target data, FU methods can be categorized into sample unlearning, client unlearning, and class unlearning \cite{liu2024survey}. 
Existing research primarily focuses on improving the efficiency of FU. For instance, FedU \cite{wang2024fedu} approximates and removes the influence of forgotten data while maintaining model utility for efficient sample unlearning. FRAMU \cite{shaik2024framu} dynamically adjusts data importance using attention mechanisms and federated reinforcement learning, thus realizing sample unlearning. FedEraser \cite{liu2021federaser} leverages historical updates retained by the central server and calibration methods to quickly eliminate client data impact, achieving client unlearning. FedRecovery \cite{zhang2023fedrecovery} computes weighted residual gradients of historical updates and applies differential privacy noise to ensure the unlearned model remains statistically indistinguishable from a fully retrained one, thereby enabling client unlearning. FURF \cite{liu2022right} employs rapid retraining, taylor expansion, and fisher information matrix approximation to efficiently erase designated client data. FUCDP \cite{wang2022federated} quantifies the contribution of CNN channels to class discrimination using a TF-IDF-based approach and prunes relevant channels to achieve class unlearning.  

\subsection{Gradient Inversion Attack in Federated Learning}
FL is designed to provide default privacy for clients by allowing multiple clients to collaboratively train a model while sharing only their local training parameters with the server without exposing their raw local data \cite{xue2024differentially}. However, recent studies have shown that the default privacy mechanism in FL is insufficient to prevent training data from being compromised by gradient reconstruction-based privacy leakage attacks. In particular, GIA can infer local training data by reconstructing gradients, thereby posing a significant threat to client privacy \cite{sun2024client}.
Existing gradient inversion attacks can be broadly classified into two classes: iterative-based attacks and recursive-based attacks \cite{ovi2023comprehensive}. Iterative-based attacks aim to minimize the distance between virtual gradients and ground-truth gradients. These methods treat the distance between gradients as the error, consider virtual inputs as optimization parameters, and formulate the recovery process as an iterative optimization problem. Representative frameworks include DLG \cite{zhu2019deep}, iDLG \cite{zhao2020idlg}, and STG \cite{yin2021see}. On the other hand, recursive-based attacks employ closed-form algorithms to reconstruct the original data. The key insight of these methods lies in leveraging the implicit relationships between input data, model parameters, and per-layer gradients to identify the optimal solution with minimal error. Examples of recursive attacks include R-GAP \cite{zhu2020r} and COPA \cite{chen2021understanding}.

\subsection{Discussion}
Existing research predominantly focuses on optimizing FU efficiency while lacking systematic and in-depth investigations into potential privacy vulnerabilities within the FU process.
Only a few studies have attempted to assess its security risks. For example, FedMUA \cite{chen2025fedmua} manipulates the FU process to alter the predictions of the global model, but its primary goal is to disrupt model performance rather than expose potential privacy leaks.  
Moreover, in the current research on machine unlearning, some researchers have proposed the machine unlearning inversion attack (MUIA) \cite{hu2024learn}, but it is not applicable to FL environments. The main reason for this is that in FL training, the interactions in each round cause the mixing of model parameters across clients. This means that the model of each client contains not only parameters from its own data but also parameters from the data of other clients. This mixing of model parameters caused by interactions creates significant challenges for inversion attacks targeting forgotten data on a specific client. Moreover, clients in FL may have highly heterogeneous data characteristics and distributions, which increases the difficulty for inversion attacks to infer specific client data from the global model. And the aggregation method in FL and other factors can also significantly affect the performance of inversion attacks.   
Therefore, to bridge this research gap, inspired by the study of GIA in FL, we propose FUIA. This is the first targeted investigation into potential privacy vulnerabilities inherent in FU.
Compared to existing research, FUIA exhibits significant differences and holds great significance for exploring more secure FU mechanisms.

\section{Preliminaries}
\label{preliminaries}
In this section, we elaborate on the general implementation of GIA within the FL framework.
Gradient inversion attacks focus on reconstructing training data through an optimization process \cite{issa2024rve}. After obtaining the gradient update \( \nabla W \), an attacker generates virtual samples \( (\hat{x}, \hat{y}) \) and feeds them into the model to compute the virtual gradient \( \nabla W' \). The attack then iteratively optimizes the virtual samples by minimizing the distance between the real gradient \( \nabla W \) and the virtual gradient \( \nabla W' \), gradually approximating the original training samples.
The optimization objective can be expressed as:
\vspace{5pt}
\begin{equation}\label{eqn-1}
    min \text{Dist} \left( \frac{\partial L(F(W, \hat{x}), \hat{y})}{\partial W}, \nabla W \right) + \text{Reg}(\hat{x}, \hat{y}),
    \vspace{5pt}
\end{equation}
where \( W \) is the current trainable parameters of the model, \( F \) is the forward propagation function of the model, \( L \) represents the loss function, $\nabla W$ is the gradient update uploaded by the client defined as:
\vspace{5pt}
\begin{equation}\label{eqn-2}
    \nabla W = \frac{\partial L(F(W, x), y)}{\partial W}.
    \vspace{5pt}
\end{equation}
And \( \text{Dist(·)} \) represents a distance function (e.g., \( l_2 \) distance or cosine similarity), commonly used to measure gradient similarity in GIA. \( \text{Reg(·)} \) represents a regularization term to ensure realistic sample reconstruction.
In image classification tasks, additional regularization techniques such as total variation \cite{rodriguez2008efficient} (to reduce noise) or clipping constraints (to limit pixel values) are often applied to produce natural-looking images. The ultimate goal is to optimize both the virtual samples \( \hat{x} \) and virtual labels \( \hat{y} \) to retrieve the local training data \( (x, y) \).

To further simplify the optimization, some studies propose label inference methods based on gradient tensor analysis \cite{wainakh2021label, yin2021see}. For instance, by analyzing the gradient distribution of the final fully connected layer, labels can be predicted before starting the optimization. This approach not only improves the quality of data reconstruction but also reduces computational complexity.

\section{Problem Statement}
\label{problem_state}

\subsection{Threat Model}
In this study, we primarily focus on classification tasks, which represent one of the most widely used applications in FL and serve as the foundation for most existing FU studies.  
We assume that the server acts as an honest-but-curious attacker, meaning that it passively records the parameter updates uploaded by each client throughout the training process and continues to track the parameter updates of target clients (i.e., those submitting unlearning requests) during the unlearning process. Apart from logging these updates, the server does not interfere with the normal training or unlearning procedures.  
This assumption is both intuitive and subtle, aligning with common FL practices and making detection difficult. In fact, many FU methods inherently rely on storing historical updates to facilitate efficient unlearning \cite{cao2023fedrecover, wang2023mitigating, liu2021federaser, yuan2023federated}.  
Once the unlearning process is complete, the server leverages the recorded model parameters to analyze differences before and after unlearning, launching an inversion attack to recover deep-level information about the forgotten data. This attack fundamentally compromises the privacy guarantees of FU.

\subsection{Problem Formulation}
In machine learning, information about the training data can be reflected through the model parameters \cite{goodfellow2016deep}. As a distributed machine learning method, FL enables each client to obtain a global model containing information from all clients through interactive training.
In FL with FedAvg as the aggregation method, the parameter update for client \(k\) in round \(t+1\) can be expressed as:
\vspace{5pt}
\begin{equation}\label{eqn-3}
    w_{(t+1)}^k = w_t - \eta \nabla L_k(w_t),
    \vspace{5pt}
\end{equation}
where \(\nabla L_k(w_t)\) is the gradient of the loss function \(L_k\) with respect to model parameters \(w_t\) on the local data. The difference between \(w_{(t+1)}^k\) and \(w_t\) arises from the gradient update using the local data. Therefore, intuitively, the difference between the two models can be approximated as the gradient information reflecting the difference in training samples.
In FU, assume that the parameter of the original global model is \(w^o\), which contains the information about the forgotten data, while the parameter of the unlearned global model is \(w^u\), which does not contain this information. Thus, the difference between the original model and the unlearned model reflects an approximation of the gradient information of the target forgotten data:
\vspace{5pt}
\begin{equation}\label{eqn-4}
    \nabla W = W^o - W^u.
    \vspace{5pt}
\end{equation}
\textbf{The first challenge} in FUIA is how to leverage the gradient information derived from this model difference to invert the forgotten data.

However, inversion based on global model differences faces inherent difficulties in FU scenarios. Specifically, the model discrepancy \(\nabla W\) essentially represents a coupling superposition of target gradients and distributed noisy gradients:
\begin{equation}
\label{eqn-5}
\nabla W \approx \eta\left(\nabla L_{\text{target}} + \sum_{i=1}^{N-1}\epsilon_i\nabla L_i\right),
\end{equation}
where \(\nabla L_{\text{target}}\) corresponds to the target gradient of the forgotten data, and \(\epsilon_i\nabla L_i\) denotes interfering gradient terms from other \(N-1\) clients. When the target forgotten data involves partial samples from a single client (with data proportion \(\rho\)), the relative strength between gradient noise and target gradient satisfies:
\vspace{5pt}
\begin{equation}
\label{eqn-7}
\frac{\|\sum_{i=1}^{N-1}\epsilon_i\nabla L_i\|}{\|\nabla L_{\text{target}}\|} \propto \frac{1-\rho}{\rho}.
\vspace{5pt}
\end{equation}
This implies that the target gradient information will be obscured by distributed noisy gradients as \(\rho \to 0\). Even when unlearning an entire client (\(\rho=1\)), the parameter aggregation mechanism specific to FL:
\vspace{5pt}
\begin{equation}
\label{eqn-7}
w^u = \sum_{k=1}^{N}\frac{n_k - \delta_k}{n_{\text{total}} - 1}w_k^u,
\vspace{5pt}
\end{equation}
where \(\delta_k \in \{0,1\}\) indicates whether client \(k\) is forgotten, still introduces gradient distortion caused by client weight adjustment into the model difference. Therefore, \textbf{the second fundamental challenge} lies in designing gradient separation algorithms to accurately reconstruct target gradient information from noisy model differences.

Besides, most methods in FU currently adopt FedAvg as the aggregation method, so FUIA is also designed based on FedAvg. This increases the difficulty of inversion attacks. In contrast, most GIA methods in FL are designed based on FedSGD, as FedSGD can directly and accurately obtain the gradient information of target data, improving the quality of reconstructed images and reducing the difficulty of attacks. However, when using FedAvg for FUIA, the gradient information obtained by the attacker is only an approximation rather than precise values, which negatively impacts the attack performance. We carry out the corresponding experimental verification and analysis in section \ref{ablationstudy}.

\section{Attack Methodology}
\label{attack_method}
\subsection{FUIA against Sample Unlearning}

\begin{figure}[h!]
    \centering
    \includegraphics[width=0.48\textwidth]{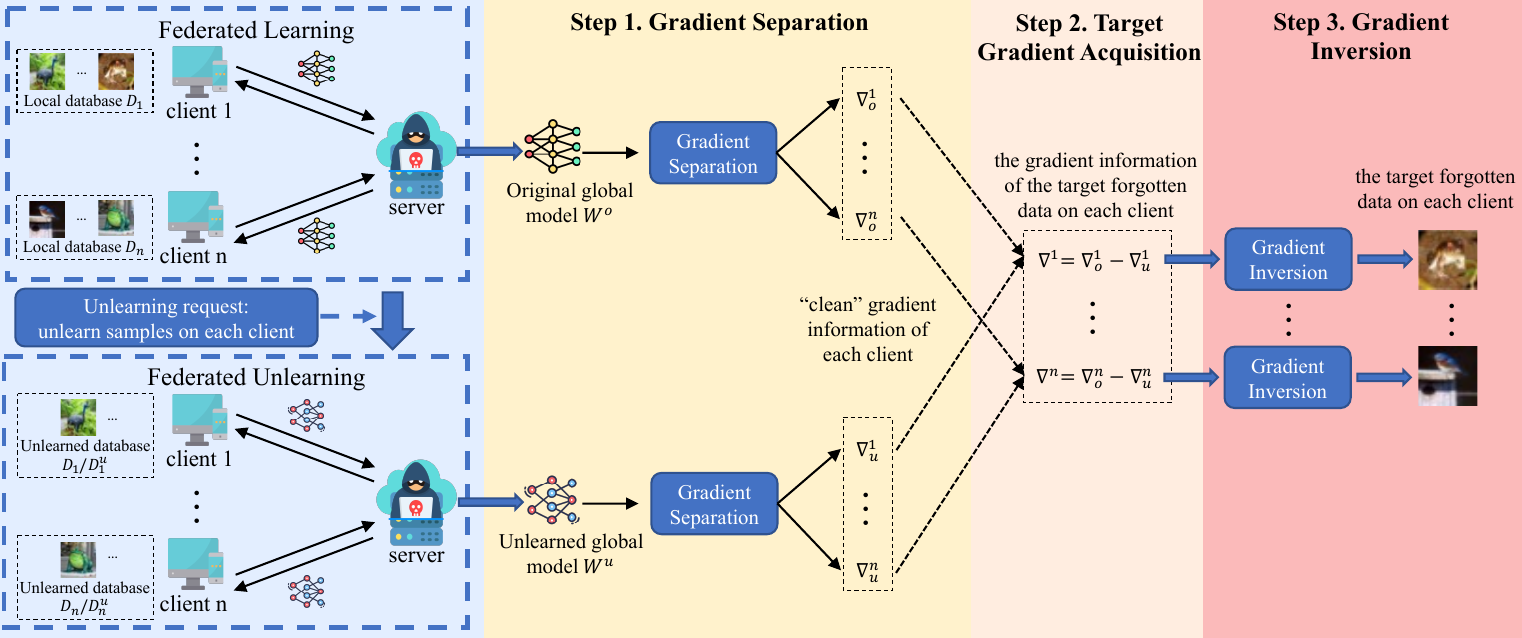} 
    \caption{Overview of FUIA against sample unlearning.}
    \label{fig_samun}
\end{figure}

\begin{algorithm}[h]
\caption{FUIA for Sample Unlearning}
\label{alg_samun}
\begin{algorithmic}[1]
\REQUIRE Original model $W^o$, unlearned model $W^u$,  unlearning request from clients $\mathcal{K}$
\ENSURE Reconstructed features $\hat{x}^*$

\FOR{each client $k \in \mathcal{K}$}
    \STATE \textbf{Step 1: Gradient Separation}\\
    \STATE \textbf{FL training phase:}
    \STATE Compute parameter differences of each round $\nabla w_t^k$ using Eq. (\ref{eqn-8}).
    \STATE Calculate $l_1$-norm sums $\text{sum}_t$ using Eq. (\ref{eqn-9}).
    \STATE Obtain clean gradients $\nabla_o^k$ using Eq. (\ref{eqn-11}).
    \STATE \textbf{FL unlearning phase:} \\
    \STATE Repeat steps 4-6 to obtain clean gradients $\nabla_u^k$.

    \STATE \textbf{Step 2: Target Gradient Acquisition}
    \STATE Derive unlearned gradients $\nabla^k = \nabla_o^k - \nabla_u^k$

    \STATE \textbf{Step 3: Gradient Inversion}
    \STATE Initialize $\hat{x}$ and get $\widehat{\nabla}$ using Eq. (\ref{eqn-14}).
    \STATE Solve optimization and get:
    \[
    \hat{x}^* = \argmin_{\hat{x}} -\frac{\langle \widehat{\nabla}, \nabla^k \rangle}{\|\widehat{\nabla}\|_2 \|\nabla^k\|_2} + \alpha \cdot TV(\hat{x})
    \]
\ENDFOR

\RETURN $\hat{x}^*$
\end{algorithmic}
\end{algorithm}

The goal of sample unlearning is to forget parts of the data from multiple clients. When these clients who wish to unlearn part of their data send an unlearn request to the server, the server and participating clients initiate the FU process and ultimately obtain the unlearned model.
For sample unlearning, the attack objective of FUIA is to invert the feature information, i.e., the image of forgotten samples. As shown in Figure \ref{fig_samun}, we divide the entire attack process into three steps: gradient separation, target gradient acquisition, and gradient inversion.
\subsubsection{Step 1: Gradient Separation}
The goal of gradient separation is to extract the ``clean" gradient information of the target client data from the global model parameters trained via FL, thereby removing the gradient noise from other clients.
Assume that client \(k\) has sent an unlearn request. Prior to unlearning, during the FL training process based on FedAvg, the parameter difference (i.e., the gradient information for the data on this client) before and after training in the \(t\)-th epoch is:
\vspace{5pt}
\begin{equation}\label{eqn-8}
    \nabla w_t^k = w_t^k - w_{(t-1)}.
    \vspace{5pt}
\end{equation}

To balance the parameter differences across all rounds and integrate the ``clean" gradient information corresponding to the local client data, the following steps are taken. 
First, we calculate the total sum of the \(l_1\) norms of parameter update differences for all participating clients in each round. Let \(C_t\) be the set of clients participating in training in the \(t\)-th epoch, then the sum of the \(l_1\) norms of parameter differences for the \(t\)-th round is:
\vspace{5pt}
\begin{equation}\label{eqn-9}
\text{sum}_t = \sum_{k \in C_t} ||\nabla w_t^k||_1.
\vspace{5pt}
\end{equation}
Next, for each client \(k\), we calculate the weight coefficient \(\gamma_t^k\) for the \(t\)-th round, which represents the relative weight of the gradient update of the client \(k\) in that round with respect to all participating clients. This coefficient is computed as:
\vspace{5pt}
\begin{equation}\label{eqn-10}
\gamma_t^k = \frac{||\nabla w_t^k||_1}{\text{sum}_t}.
\vspace{5pt}
\end{equation}
And last, by weighted averaging the gradient differences \(\nabla w_t^k\) for each round, we can compute the ``clean" gradient information for client \(k\), denoted \(\nabla_o^k\):
\vspace{5pt}
\begin{equation}\label{eqn-11}
\nabla_o^k = \sum_{t=1}^{E} \gamma_t^k \cdot \nabla w_t^k,
\vspace{5pt}
\end{equation}
where \(E\) is the total number of training rounds, and \(\nabla_o^k\) represents the ``clean" gradient information for client \(k\), which has removed noise from other clients.
Similarly, using the stored parameter updates during the FU process, we can obtain the ``clean" gradient information for client \(k\) after removing the forgotten data, denoted \(\nabla_u^k\).

\subsubsection{Step 2: Target Gradient Acquisition}
The goal of this step is to use the difference between the ``clean" gradient information before and after unlearning to acquire the gradient information of the target forgotten data.
From the gradient separation step, we obtain the ``clean" gradient information \(\nabla_o^k\) for client \(k\) during FL training, which contains the gradient information for all the data on the client. After unlearning, the ``clean" gradient information for client \(k\) is \(\nabla_u^k\), which contains only the gradients for the remaining data after removing the forgotten data.
Thus, the gradient information for the target forgotten data on client \(k\) is:
\vspace{5pt}
\begin{equation}\label{eqn-12}
\nabla^k = \nabla_o^k - \nabla_u^k.
\vspace{5pt}
\end{equation}

\subsubsection{Step 3: Gradient Inversion}
Following the previous steps, we extract the gradient information of the target forgotten data from the global model parameters while effectively removing noise from other clients. 
Inspired by GIA in FL, the goal of gradient inversion is to decode the gradient information of forgotten data using optimization algorithms to recover its features. 
However, not all GIA optimization algorithms are suitable for FUIA. The primary reason is that, to ensure applicability in a broader range of scenarios and remain inconspicuous, we assume the server, as an attacker, operates under a relatively relaxed setting, behaving as an honest-but-curious attacker rather than a malicious one. Besides, we assume the server does not have explicit knowledge of the FU mechanism. Consequently, the server only obtains an approximate gradient of the forgotten data rather than its exact value. Additionally, some GIA methods require auxiliary information beyond gradients, such as data distribution, to facilitate reconstruction. However, we assume the attacker strictly adheres to the FL protocol and cannot access any additional information beyond the gradients.

To address these challenges, we adopt the optimization algorithm from  \cite{geiping2020inverting} to recover the feature information of the forgotten data based on its gradient information. The principle of GIA is discussed in Section \ref{preliminaries}. Specifically, we choose cosine similarity as the distance function to measure gradient similarity, focusing more on the direction of the gradients because it contains crucial information on how the model responds to input changes. Besides, since we have obtained an approximation of the gradient for the forgotten data instead of exact values, we focus more on the direction of the gradients rather than their magnitude. We also include total variation (TV) as a regularization term to promote smoothness in the reconstructed image while preserving edge information.
Based on the steps above, we obtain the gradient information \(\nabla^k\) of the target forgotten data on client \(k\). Therefore, the objective of the optimization algorithm can be expressed as:
\vspace{5pt}
\begin{equation}\label{eqn-13}
    min-\frac{<\widehat{\nabla},\nabla^k>}{\left\|\widehat{\nabla}\right\|_2\left\|\nabla^k\right\|_2}+\alpha \cdot TV(\widehat{x}),
    \vspace{5pt}
\end{equation}
where $\widehat{\nabla}$ represents the gradient computed based on model parameters \(W^o\) and the virtual sample \(\hat{x}\) defined as:
\vspace{5pt}
\begin{equation}\label{eqn-14}
    \widehat{\nabla}=\frac{\partial L(F(W^o,\hat{x}),\hat{y})}{\partial W^o}.
    \vspace{5pt}
\end{equation}
The total variation regularization term \(TV(\hat{x})\) is defined as: 
\vspace{5pt}
\begin{equation}
\label{eqn-15}
TV(\hat{x}) = \sum_{i,j} \sqrt{(x_{i+1,j} - x_{i,j})^2 + (x_{i,j+1} - x_{i,j})^2}.
\vspace{5pt}
\end{equation}   
This term helps suppress noise in the reconstructed image while preserving edge details. \(\alpha\) is a hyperparameter balancing the two terms.
The goal is to use the optimization algorithm to make the gradient of the constructed virtual sample as close as possible to the gradient of the target forgotten data, thereby obtaining a virtual sample that closely matches the target forgotten data. Additionally, during the optimization process, we assume that the data labels are known. As mentioned earlier, previous studies have shown that data labels can be directly extracted in GIA \cite{yin2021see, wainakh2021label}.

\subsection{FUIA against Client Unlearning}

\begin{figure}[h!]
    \centering
    \includegraphics[width=0.45\textwidth]{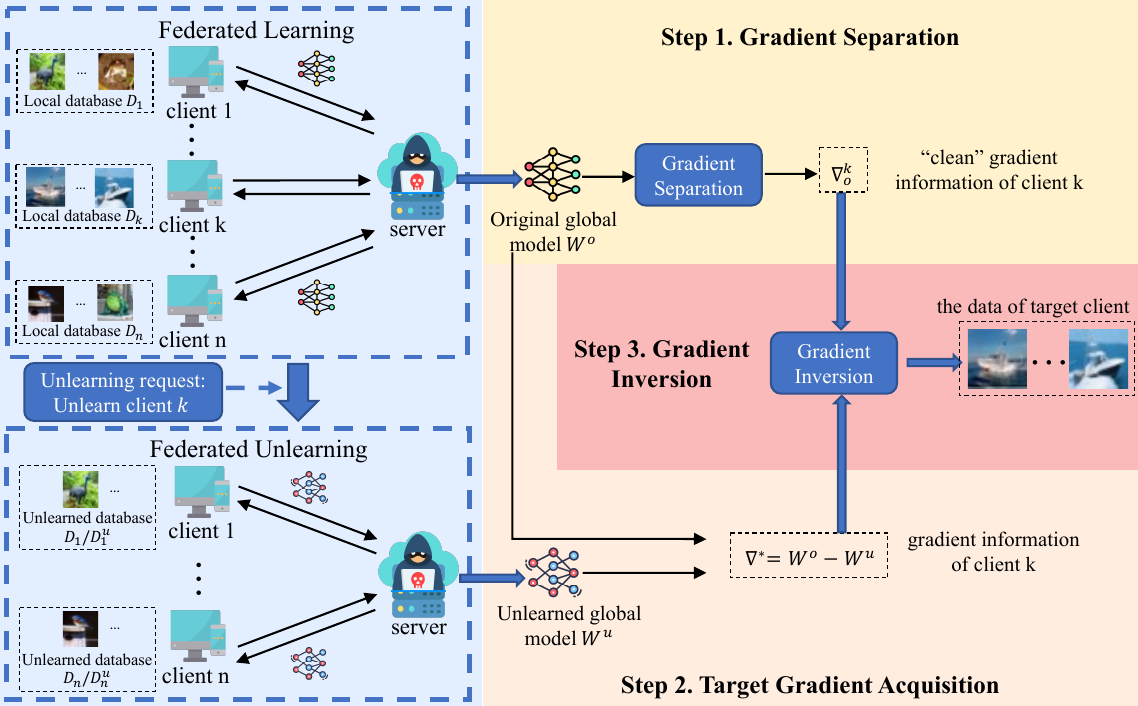} 
    \caption{Overview of FUIA against client unlearning.}
    \label{fig_cliun}
\end{figure}

\begin{algorithm}[h]
    \caption{FUIA for Client Unlearning}
    \label{alg_cliun}
    \begin{algorithmic}[1]
    \REQUIRE Original model $W^o$, unlearned model $W^u$,  unlearning request from clients $\mathcal{K}$
    \ENSURE Reconstructed feature information of clients $\mathcal{K}$, denoted as $\hat{x}^*$.
        \FOR{each client $k \in \mathcal{K}$}
        \STATE \textbf{Step 1: Gradient Separation}\\
        \STATE \textbf{FL training phase:}
        \STATE Similar to in sample unlearning, Obtain clean gradients $\nabla_o^k$ using Eq. (\ref{eqn-16}).
    
        \STATE \textbf{Step 2: Target Gradient Acquisition}
        \STATE Derive unlearned gradients $\nabla^* = W^O - W^*$.
    
        \STATE \textbf{Step 3: Gradient Inversion}
        \STATE Initialize $\hat{x}$ and get $\widehat{\nabla}$.
        \STATE Solve optimization and get $\hat{x}^*$ by:
            \[
            min - \left( (1 - \gamma) \cdot l(\widehat{\nabla}, \nabla_o^k) + \gamma \cdot l(\widehat{\nabla}, \nabla^*) \right) + \alpha \cdot TV(\hat{x}).
            \]
    \ENDFOR
\RETURN $\hat{x}^*$.
\end{algorithmic}
\end{algorithm}

In the context of client unlearning, the goal is to unlearn all the data from a specific client. After the client sends an unlearning request, the server and client initiate the unlearning process, ultimately obtaining a global model that no longer includes the data of the target client. In this scenario, the goal of the FUIA is to revert the feature information of the target client. Similar to sample unlearning, the entire attack process can be divided into three steps: gradient separation, target gradient acquisition, and gradient inversion. However, compared to sample unlearning, the implementation of the attack process in client unlearning differs in the specific operations of each step. The overview of FUIA against client unlearning is shown in Figure \ref{fig_cliun}.
\subsubsection{Step 1: Gradient Separation}
In the gradient separation step, the objective is to extract the “clean” gradient information of the data on the target client from the global model parameters obtained through FL. Suppose that client $k$ makes the unlearning request, and after the FU process, the data on client $k$ has been entirely unlearned, meaning that the global model no longer contains their gradient information. Therefore, unlike sample unlearning, this step does not require focusing on the gradient changes in the unlearning process but only needs to separate the gradient information of client $k$ from the FL training process. Specifically, by calculating the parameter differences for each round and weighting them, we can obtain the “clean” gradient information of client $k$, denoted as \(\nabla_o^k\), as shown in the following formula:
\vspace{5pt}
\begin{equation}\label{eqn-16}
\nabla_o^k = \sum_{t=1}^{E} \gamma_t^k \cdot \nabla w_t^k,
\vspace{5pt}
\end{equation}
where \(E\) is the total number of training epochs, and \(\nabla_o^k\) is the “clean” gradient information of client $k$, with noise from other clients removed.

\subsubsection{Step 2: Target Gradient Acquisition}
We leverage the “clean” gradient information (\(\nabla_o^k\)) obtained in the previous step, along with the gradient information difference between the global model before and after unlearning (\(\nabla^*\)), to further refine the gradients of the data on the target client. Let the global model parameters before unlearning be \(W^o\), and after unlearning, they are \(W^u\), then the gradient information for the target client is the difference between these two, expressed as:
\vspace{5pt}
\begin{equation}\label{eqn-17}
\nabla^* = W^o - W^u.
\vspace{5pt}
\end{equation}
The goal of this step is to enhance the gradient description of the target client by supplementing it with the gradient information from the global model difference. However, due to parameter interactions in the FL training process, \(\nabla^*\) contains noise from other clients, and hence, we treat it as auxiliary information rather than directly using it.

\subsubsection{Step 3: Gradient Inversion}
Finally, in the gradient inversion step, the goal of the optimization algorithm is to use the gradient information obtained in the previous step to reveal the feature information of the data on the target client. Similar to sample unlearning, this process is also implemented through the optimization algorithm, with the objective function given by:
\vspace{5pt}
\begin{equation}\label{eqn-18}
min -((1-\gamma)*l(\widehat{\nabla},\nabla_o^k)+\gamma*l(\widehat{\nabla},\nabla^*))+\alpha TV(\widehat{x}).
\vspace{5pt}
\end{equation}
$l(\widehat{\nabla},\nabla^*)$ is the cosine similarity between the gradient vectors defined as:
\vspace{5pt}
\begin{equation}\label{eqn-19}
    l(\widehat{\nabla},\nabla^*)=\frac{<\widehat{\nabla},\nabla^*>}{\left\|\widehat{\nabla}\right\|_2\left\|\nabla^*\right\|_2}.
\vspace{5pt}
\end{equation}
$\widehat{\nabla}$ represents the gradient of the constructed virtual sample \((\hat{x}, \hat{y})\), \(\text{TV}(x^{'})\) is the total variation regularization term used to control the smoothness of the reconstructed image and reduce noise and artifacts, and \(\alpha\) is a hyperparameter that balances the regularization term. 
To balance the influence of \(\nabla^k\) and \(\nabla^*\), we set \(\gamma = 0.1\) and analyze it in detail in section \ref{ablationstudy}. 
Through this optimization process, the attacker can construct virtual samples whose features closely resemble the data on the target client, successfully revealing their feature information.

\subsection{FUIA against Class Unlearning}

\begin{figure}[h!]
    \centering
    \includegraphics[width=0.45\textwidth]{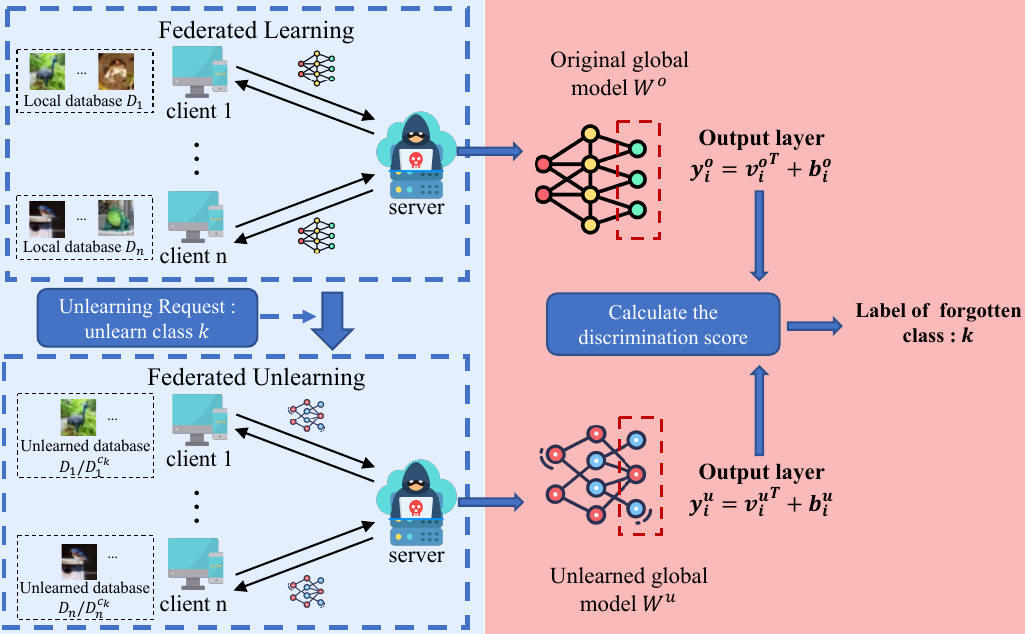} 
    \caption{Overview of FUIA against class unlearning.}
    \label{fig_claun}
\end{figure}

\begin{algorithm}[h]
\caption{FUIA for Class Unlearning}
\label{alg_claun}
\begin{algorithmic}[1]
    \REQUIRE Original model $W^o$, unlearned model $W^u$.
    \ENSURE The label of the forgotten class.
    \STATE Obtain $b_o$ and $v_o$ from original model $W^o$, and obtain $b_u$ and $v_u$ from unlearned model $W^u$.
    \FOR{each class $i$ in parallel:}
        \STATE Calculate weight and bias differences for each class using Eq. (\ref{eqn-20}) and get $v_{diff}[i]$ and $b_{diff}[i]$.
        \STATE Calculate the discrimination score $S_d[i]$ by:\\ 
        $S_d[i] = \beta \cdot \frac{v_{\text{diff}}[i]}{\sum_{i=1}^n v_{\text{diff}}[i]} + (1-\beta) \cdot \frac{b_{\text{diff}}[i]}{\sum_{i=1}^n b_{\text{diff}}[i]}.$
    \ENDFOR
    \STATE Identify the class with the maximum discrimination score:\\
    $\text{$class\_id$} = \arg \max_i (\text{$S_d$}[i]).$
\RETURN The label of the forgotten class: \text{$class\_id$}.
\end{algorithmic}
\end{algorithm}

In the context of class unlearning, the goal is to unlearn all data of certain classes. Unlike client unlearning or sample unlearning, the core of class unlearning lies in forgetting data from specific classes, which may be distributed across multiple clients. Since we make no specific assumptions about the data distribution in FL, and the data is non-iid (non-independent and identically distributed), data from a given class may randomly appear across multiple clients. For the server acting as an attacker, although it cannot directly access data from individual clients, it has knowledge of the entire training dataset. Therefore, in class unlearning, the objective of the FUIA is to infer the labels of the target forgotten class. By leaking these labels, the attacker can obtain sensitive privacy information regarding the training data. The overview of FUIA against class unlearning is shown in Figure \ref{fig_claun}.

Existing studies have shown that the removal of large amounts of data has a significant impact on the model parameters \cite{goodfellow2016deep}. The training process of deep learning models essentially optimizes the model parameters to improve performance. When data from a specific class is deleted, the model parameters, particularly those in the output layer associated with that class, will change \cite{zhang2022all}. This is because, after deleting the data points, the model no longer receives training signals from these data, which causes the weights of the output nodes related to that class to update less frequently and may even diminish or converge to zero. After the unlearning process removes data from a specific class, the server, acting as the attacker, can infer the characteristics of that class by comparing the weight differences in the output layer before and after the unlearning.

For a multiclass classification problem, suppose there are \(n\) classes, and the score \(y_i\) for each class \(i\) can be expressed as: \(y_i = v_i^T + b_i\), where \(v_i\) is the weight vector corresponding to class \(i\) (size \(m \times n\)), and \(b_i\) is the bias term for class \(i\) (size \(n\)). We define a discrimination score $S_d$ to represent the difference in the output layer before and after unlearning. The class with the highest score corresponds to the forgotten class.
Assuming the weight matrix and bias of the output layer before unlearning are \(v_o\) and \(b_o\), and \(v_u\) and \(b_u\) after unlearning, we can use the $l_1$ norm to measure the changes. The weight matrix difference and bias difference are computed as:
\begin{equation}\label{eqn-20}
\begin{split}
    v_{\text{diff}}[i] &= \sum_{j=1}^m ||v_o[i,j] - v_u[i,j]||_1, \\
    b_{\text{diff}}[i] &= ||b_o[i] - b_u[i]||_1.
\end{split}
\end{equation}
Therefore, the discrimination score for each class can be defined as:
\vspace{5pt}
\begin{equation}\label{eqn-21}
\text{$S_d$}[i] = \beta \cdot \frac{v_{\text{diff}}[i]}{\sum_{i=1}^n v_{\text{diff}}[i]} + (1-\beta) \cdot \frac{b_{\text{diff}}[i]}{\sum_{i=1}^n b_{\text{diff}}[i]},
\vspace{5pt}
\end{equation}
where we set \(\beta=0.5\) to equally balance the contributions of the weight and bias differences. Based on the discrimination score, we can determine the forgotten class. The class with the highest score is considered the unlearned class. The class id for unlearning can be determined as:
\vspace{5pt}
\begin{equation}\label{eqn-22}
\text{$class\_id$} = \arg \max_i (\text{$S_d$}[i]).
\vspace{5pt}
\end{equation}
In cases where \(k > 1\) classes have been unlearned, we sort the discriminative scores \(\text{diff\_score}\) and select the top-\(k\) scores to obtain the labels of forgotten classes. 
Through this process, the attacker can accurately identify the forgotten class, thereby leaking privacy information about the training data.

\section{Experiments and Analysis}
\label{experiments}
\subsection{Experimental Settings}
\subsubsection{Datasets}
We select CIFAR-10 and CIFAR-100 \cite{krizhevsky2009learning} to evaluate the proposed FUIA. These datasets are widely used in image classification tasks and have been extensively utilized in existing FU research. CIFAR-10 contains 60,000 32x32 pixel color images across 10 classes, suitable for basic classification tasks. CIFAR-100, in contrast, includes 100 classes, making the task more challenging and suitable for evaluating more complex classification models.

In the experiments, we employed pre-training in the FL training process for sample unlearning and client unlearning tasks. Specifically, we randomly partitioned the dataset into two disjoint subsets, with 80$\%$ used to train a pre-trained model and the remaining 20$\%$ serving as private data for individual clients. After pre-training, further training was conducted on client-specific data. This approach is widely adopted in FL as it enhances model performance. By leveraging large-scale data to capture generalized features, the pre-trained model provides a more stable initialization for client fine-tuning, thereby reducing training time and computational overhead while improving convergence speed and generalization capability \cite{tian2022fedbert}.
In the class unlearning task, we do not adopt a pre-training approach. This decision stems from the fact that data from the target class to be unlearned may exist in both the training data of the pre-trained model and the private datasets of the clients. Since our dataset partitioning is random and does not enforce a specific class to be client-private, it contradicts the fundamental assumption of pre-training, which assumes that class data can be shared between training and testing datasets. 

\subsubsection{Models}
In the experiments, we select two different model architectures to evaluate the performance of FUIA on different datasets. For the CIFAR-10 dataset, we use ResNet-18 \cite{he2016deep}, a relatively small network suitable for simpler image classification tasks. For the CIFAR-100 dataset, we employ ResNet-44 \cite{he2016deep}, a deeper architecture that is better suited for handling more complex data with a larger number of classes. 

\subsubsection{FL settings}
We configure 10 clients, with 50$\%$ randomly selected to participate in each training round. Each participating client performs 3 local training rounds on its own data. In the standard experiment, we use FedAvg as the aggregation method, which is widely applied in FL and FU.

\subsubsection{FU settings}

To comprehensively analyze the privacy vulnerabilities of FU, we select two representative unlearning methods for each of the three FU scenarios to evaluate the effectiveness of FUIA. Retraining, as the most fundamental method, applies to all FU scenarios and represents exact FU methods. Additionally, we selected widely studied representatives of approximate FU for testing. For class unlearning, we chose FUCDP \cite{wang2022federated} for evaluation; for client unlearning, we use FUEraser \cite{liu2021federaser} as a comparative method; and for sample unlearning, we perform unlearning on the target client using UnrollingSGD \cite{thudi2022unrolling} and conduct global model aggregation through the standard FL process.

\subsubsection{Metrics}
In the FUIA for sample unlearning and client unlearning, we use \textbf{MSE} (Mean Squared Error) and \textbf{PSNR} (Peak Signal-to-Noise Ratio) to evaluate the effectiveness of image reconstruction \cite{yue2023gradient}. Specifically, MSE reflects the difference between the original image and the reconstructed image, with a smaller difference indicating a higher similarity between the two, while PSNR measures the signal-to-noise ratio of the reconstructed image, with a higher ratio indicating better image quality and more effective attacks. For class unlearning, we use prediction accuracy to assess the attack effectiveness. Specifically, we evaluate whether the predicted labels match the actual labels of the forgotten classes.

\subsubsection{Comparison Method and Implementation}
We adopt the MUIA \cite{hu2024learn} and apply it within the FL environment for comparative analysis. Specifically, we utilize the difference between the global models of FL before and after unlearning as the gradient information of the forgetting data for MUIA. All experiments are performed on a computing system equipped with an NVIDIA 3090 GPU (24 GB of VRAM) and Intel Xeon processors.

\subsection{Results Analysis}

\subsubsection{Effectiveness of FUIA against Sample Unlearning}
\begin{figure}[h!]
    \centering
    \includegraphics[width=0.45\textwidth]{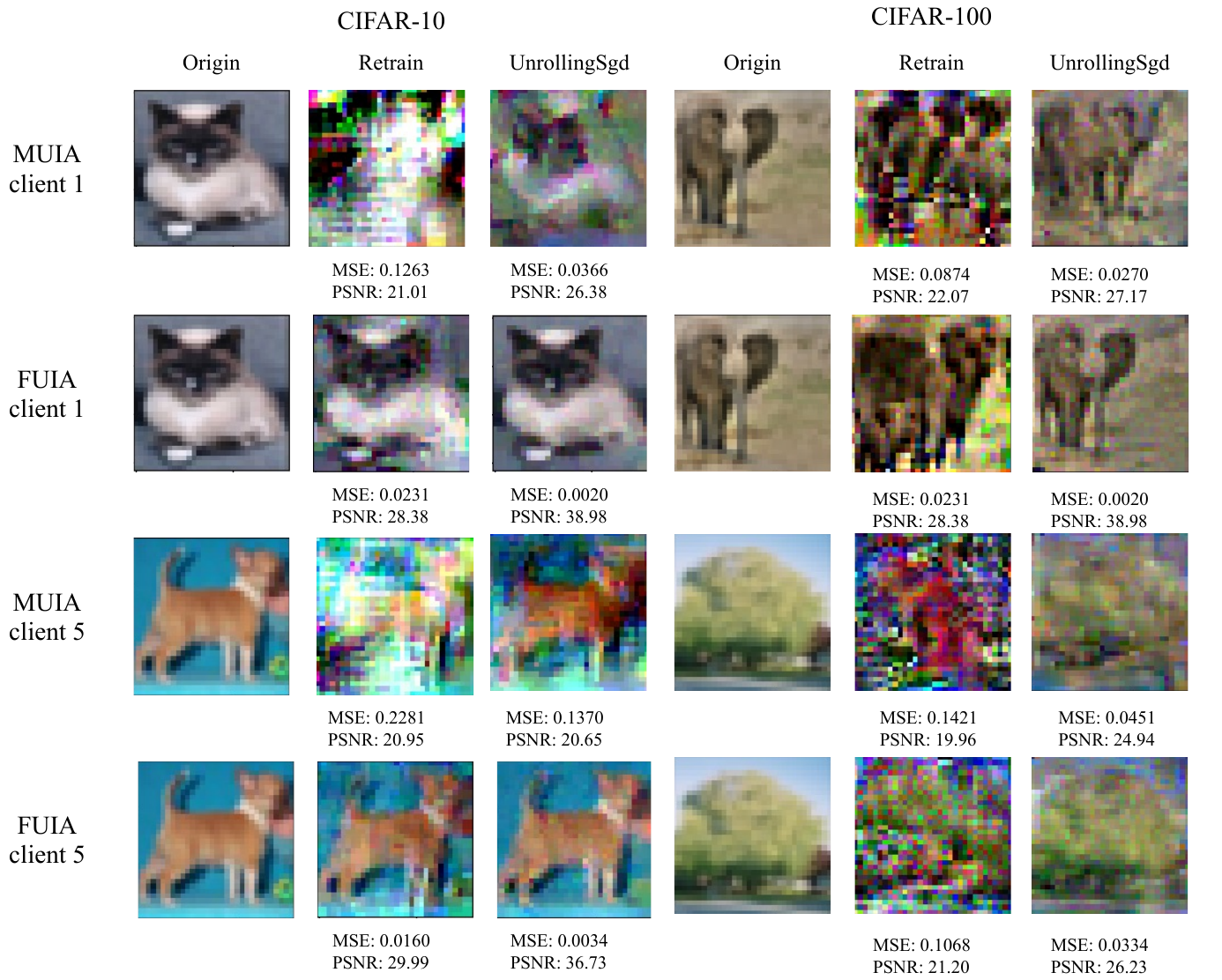} 
    \caption{Effectiveness of FUIA against sample unlearning.}
    \label{fig_ex_samun}
\end{figure}

To conveniently demonstrate the effectiveness of the attack, we start by exploring a relatively simple scenario. Specifically, to reduce computational cost and improve experimental efficiency, we set the sample size on each client to 8 and randomly select 1 data point for unlearning. In Section \ref{ablationstudy}, we conduct additional experimental analysis regarding the number of samples per client and the number of forgotten data to explore how these factors influence the attack performance.
The results are shown in Figure \ref{fig_ex_samun}, where we randomly select the reconstruction results of two clients for display. The figure illustrates the reconstruction performance of FUIA and MUIA against two FU methods, Retraining and UnrollingSgd, on the CIFAR-10 and CIFAR-100 datasets. To quantitatively evaluate the reconstruction performance, we also analyzed the results using MSE and PSNR.

Specifically, FUIA successfully reconstructs the forgotten data for both Retraining and UnrollingSgd, with reconstructed images that leak a significant amount of private information. Both the visual quality of the images and the values of metrics such as MSE and PSNR show that FUIA performs exceptionally well. In contrast, the reconstruction results of MUIA are noticeably inferior, especially in terms of image clarity and detail. This is mainly due to the coarse gradient information obtained by MUIA regarding the forgotten data, which includes more noise and fails to capture the fine details of the forgotten data.
Furthermore, the reconstruction performance of FUIA for the Retraining method is still somewhat inferior compared to UnrollingSgd. We attribute this discrepancy to the fundamental differences in unlearning mechanisms. As a strict exact FU method, Retraining provides FUIA with limited and less precise gradient information on forgotten data. In contrast, UnrollingSGD, as an approximate FU method, offers richer and clearer gradient information, thereby enhancing the reconstruction capability of FUIA. However, Retraining, being the most naïve unlearning approach, is rarely practical, while approximate FU methods such as UnrollingSGD dominate federated unlearning research. In summary, FUIA demonstrates high effectiveness in attacking sample unlearning methods, successfully reconstructing relatively clear images under the current experimental settings, leading to severe privacy leakage and significantly outperforming MUIA.

\subsubsection{Effectiveness of FUIA against Client Unlearning}
\begin{figure}[h!]
    \centering
    \includegraphics[width=0.45\textwidth]{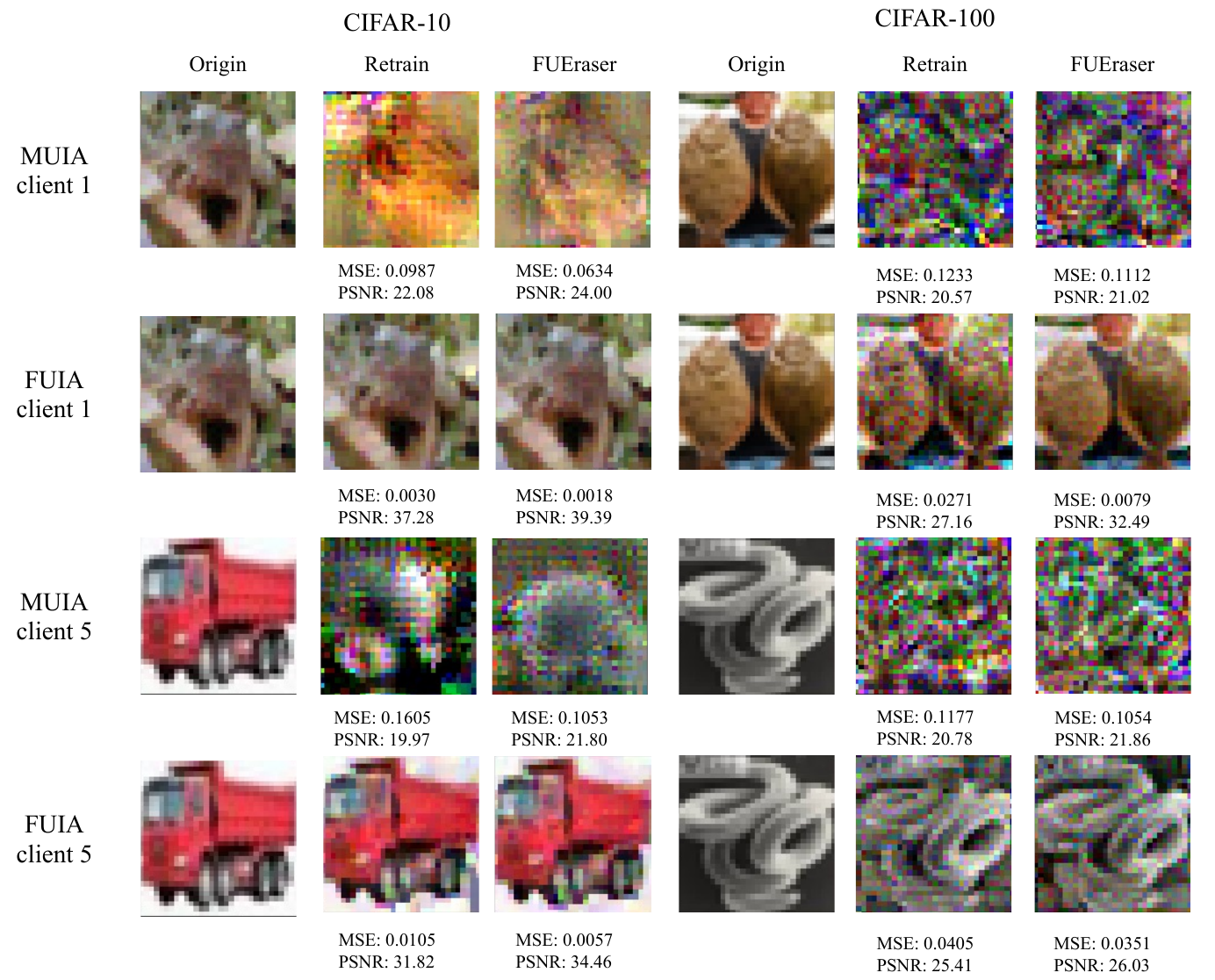} 
    \caption{Effectiveness of FUIA against client unlearning.}
    \label{fig_ex_cliun}
\end{figure}

Similarly, for client unlearning, we also begin the experiments with a simple scenario. Specifically, we set the number of data points per client to 1, and the unlearning target is to randomly select 1 client and unlearn all its data. While this setup is simple, it effectively demonstrates the attack performance on client data unlearning. In Section \ref{ablationstudy}, we conduct additional experimental analysis on the number of data points per client to explore how different data quantities affect the attack performance.
To demonstrate the performance of FUIA under different conditions, we randomly selected the results of two experiments for display, focusing on the reconstruction effects of FUIA against two FU methods, Retraining and FUEraser, on the CIFAR-10 and CIFAR-100 datasets. Additionally, we performed a quantitative evaluation of the reconstruction quality using MSE and PSNR.

AS shown in Figure \ref{fig_ex_cliun}, FUIA successfully reconstructs the data of the target client for both Retraining and FUEraser, indicating that client unlearning-based FU methods can lead to privacy leakage. Both in terms of image detail recovery and performance in metrics such as MSE and PSNR, FUIA shows significant effectiveness. Similar to the results for sample unlearning, the reconstruction performance of FUIA is clearly superior to that of MUIA, particularly in terms of image clarity and detail recovery.
However, FUIA still shows some performance gaps when applied to Retraining compared to approximate FU methods like FUEraser.
In conclusion, FUIA is highly effective in client unlearning attacks, capable of reconstructing relatively clear images under the current setup, leading to significant privacy leakage in FU scenarios. Furthermore, it outperforms MUIA.

\subsubsection{Effectiveness of FUIA against Class Unlearning}
\begin{figure}[h!]
    \centering
    \begin{subfigure}[b]{0.24\textwidth}
        \centering
        \includegraphics[width=\textwidth]{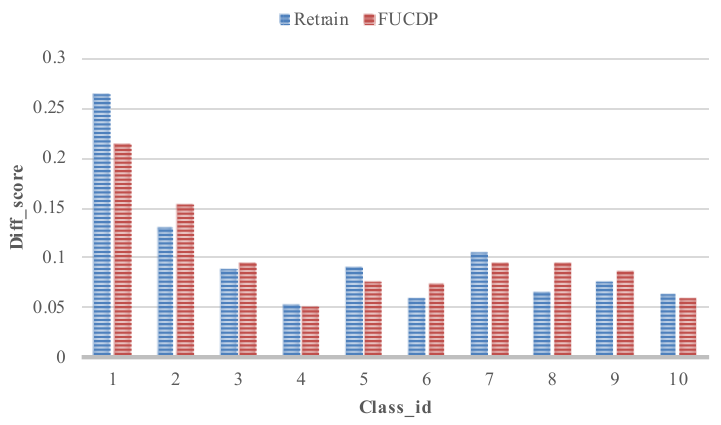} 
        \caption{Unlearning class 0}
        \label{fig_ex_claun_class0}
    \end{subfigure}
    \hfill
    \begin{subfigure}[b]{0.24\textwidth}
        \centering
        \includegraphics[width=\textwidth]{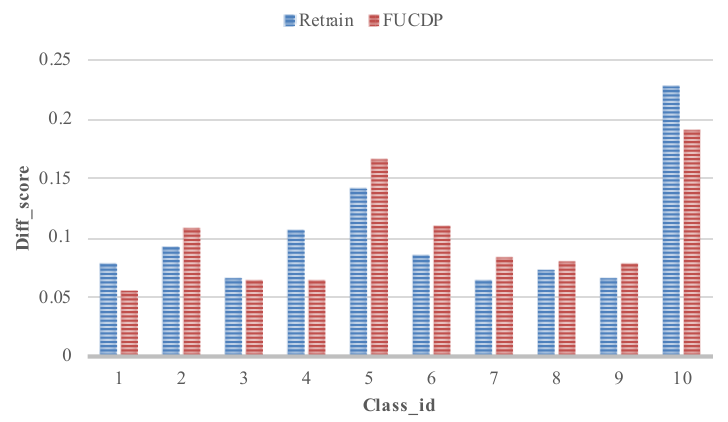} 
        \caption{Unlearning class 9}
        \label{fig_ex_claun_class9}
    \end{subfigure}
    \caption{Effectiveness of FUIA against class unlearning.}
    \label{fig_ex_claun_class0_and_class9}
\end{figure}

\begin{table}[h]
    \caption{The Predictive probability of FUIA against class unlearning. $N_{uc}$ represents the number of forgotten classes. $N_{ic}$ represents the number of correct classes inferred by FUIA.}
    \label{table_claun}
    \centering
    \setlength{\tabcolsep}{6mm}
    \renewcommand{\arraystretch}{1.25}
    \begin{tabular}{ccc}
        \toprule[1.5pt]
        $N_{uc}$ & $N_{ic}$ & Predictive Probability \\ \hline
        \multirow{2}{*}{1} & 0 & 0$\%$ \\
         & 1 & 100$\%$ \\ \hline
        \multirow{2}{*}{2} & 0,1 & 0$\%$ \\
         & 2 & 100$\%$ \\ \hline
        \multirow{2}{*}{3} & 0,1,2 & 0$\%$ \\
         & 3 & 100$\%$ \\ \hline
        \multirow{3}{*}{4} & 0,1,2 & 0$\%$ \\
         & 3 & 20$\%$ \\
         & 4 & 80$\%$ \\
         \bottomrule[1.5pt]
    \end{tabular}
\end{table}
To better illustrate the effectiveness of the attack, we primarily conduct experiments using the CIFAR-10 dataset. In each experiment, we randomly select data from one class out of ten classes for unlearning and use FUIA to analyze the differences between the models before and after unlearning. By calculating the list of discrimination scores, we infer the labels of the unlearned class data. Furthermore, in Section \ref{ablationstudy}, we conduct additional experiments on unlearning multiple classes to verify the broad applicability and robustness of FUIA.

The experimental results are shown in Figure \ref{fig_ex_claun_class0_and_class9}, where we randomly select two experimental instances to demonstrate the attack effectiveness of FUIA under the Retraining and FUCDP unlearning methods. The figure presents the discrimination scores for each data class, with red and blue bars representing the Retraining and FUCDP methods, respectively. In Figure \ref{fig_ex_claun_class0}, when unlearning data from class 0, FUIA accurately identifies class 0 as having the highest discrimination score, regardless of the unlearning method used. Similarly, in Figure \ref{fig_ex_claun_class9}, when unlearning class 9, FUIA consistently reveals class 9 as having the highest discrimination score, demonstrating the robustness of the attack.  
Furthermore, we measure the probability of FUIA correctly inferring the forgotten class, as shown in Table \ref{table_claun}. When unlearning a single class, FUIA achieves a 100$\%$ success rate in identifying the forgotten class. These results indicate that FUIA directly infers the labels of unlearned data, significantly exacerbating privacy leakage risks.

\section{Ablation Study}
\label{ablationstudy}
\subsection{Ablation Study for FUIA against Sample Unlearning}
\subsubsection{The Number of Data on Each Client}
\begin{figure}[h!]
    \centering
    \includegraphics[width=0.45\textwidth]{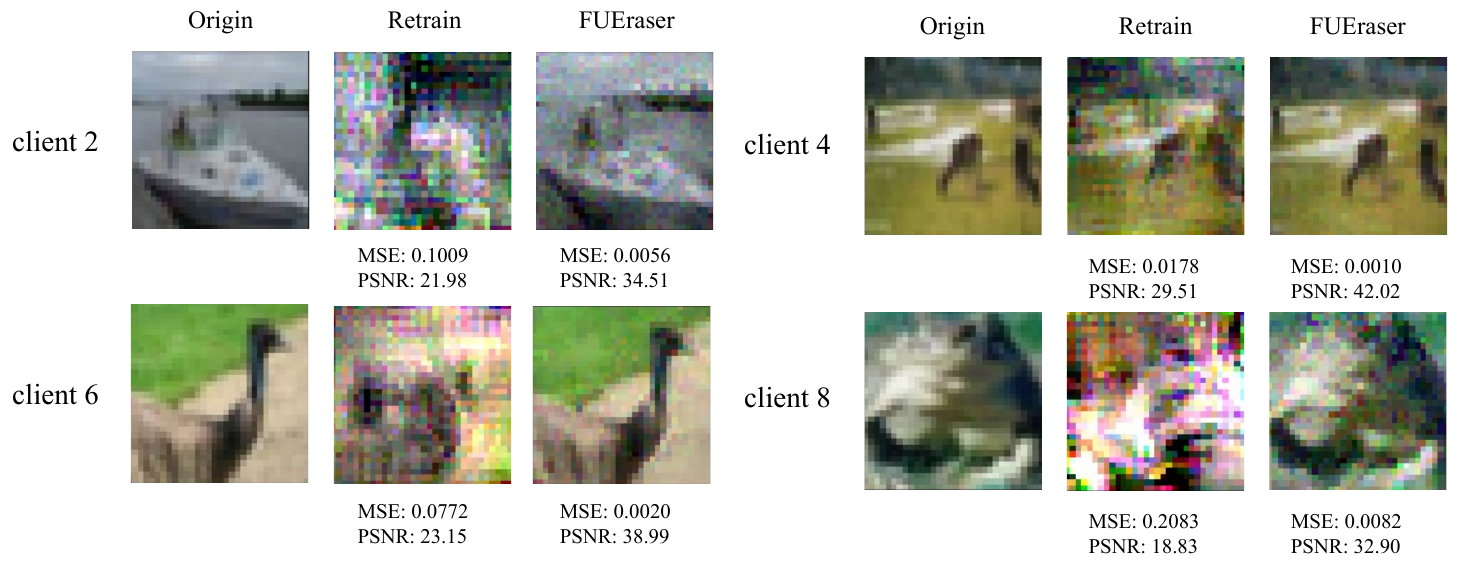} 
    \caption{Effectiveness of FUIA against sample unlearning with different numbers of client data.}
    \label{fig_ex_samun_abl_128un1}
\end{figure}

To explore a more generalizable scenario, we further analyze the impact of client data volume on experimental results. Specifically, we use CIFAR-10 as the experimental dataset, expand the data size of each client to 256, randomly select one data sample for unlearning, and then apply FUIA to attack the forgotten data. As shown in Figure \ref{fig_ex_samun_abl_128un1}, we randomly select four clients to present their reconstruction results. The results indicate that, compared to clients with smaller data volumes, FUIA maintains strong attack performance against UnrollingSgd, an approximate FU method, producing relatively clear reconstructed images with MSE and PSNR values comparable to previous experiments. However, when attacking Retraining, an exact FU method, the attack performance of FUIA deteriorates. Only one client can be reconstructed with clarity, while the remaining reconstructions are blurry, with significantly increased MSE and relatively lower PSNR values.  
We hypothesize that this phenomenon arises because the increase in data volume leads to more complex and diverse feature learning during client training. As a result, the gradient information obtained by FUIA for forgotten data becomes more ambiguous and uncertain, affecting reconstruction accuracy and degrading image quality.  
Overall, client data volume has some impact on the attack performance of FUIA. As data volume increases, its effectiveness in attacking exact FU methods diminishes. However, for approximate FU methods, FUIA remains highly effective in reconstructing forgotten data.

\subsubsection{The Number of Forgotten Data on Each Client}
\begin{figure}[h!]
    \centering
    \includegraphics[width=0.45\textwidth]{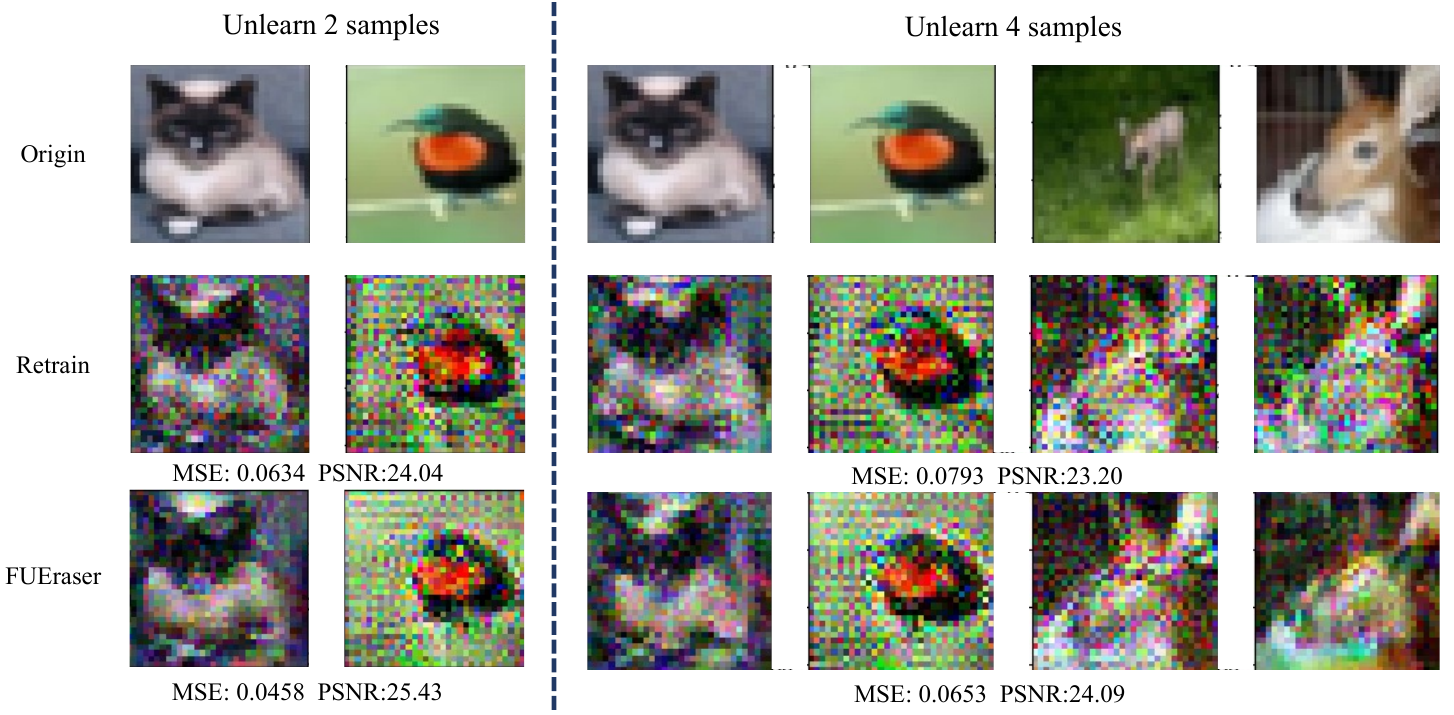} 
    \caption{Effectiveness of FUIA against sample unlearning with different numbers of forgotten data.}
    \label{fig_ex_samun_abl_8un2_4}
\end{figure}

We further investigate the impact of the number of forgotten data samples per client on the effectiveness of FUIA attacks. Specifically, we use the CIFAR-10 dataset and select 2 and 4 data samples for unlearning on each client, followed by an attack using FUIA. The experimental results are shown in Figure \ref{fig_ex_samun_abl_8un2_4}, where we present the attack outcomes for one randomly selected client. When each client forgets 2 data samples, FUIA remains effective in reconstructing the images, preserving certain original data features and causing partial privacy leakage. However, compared to the scenario where only 1 data sample is forgotten, the reconstructed images exhibit reduced clarity, with less distinct details and noticeable parameter deviations. When the number of forgotten samples increases to 4 per client, the attack performance of FUIA deteriorates further. While some reconstructed images still retain certain features of the forgotten data, they become significantly more blurred, and in some cases, it becomes impossible to fully reconstruct all forgotten samples, with some feature information completely lost.  
We assume that this phenomenon is primarily due to the increasing difference in model parameters before and after unlearning as the number of forgotten data samples grows. Notably, since FUIA relies on estimated rather than exact gradient information, larger parameter variations lead to more ambiguous target gradients, thereby weakening the attack performance. In conclusion, increasing the number of forgotten data samples per client reduces the effectiveness of FUIA.

\subsection{Ablation Study for FUIA against Client Unlearning}
\subsubsection{The value of \(\gamma\)}
\begin{figure}[h!]
    \centering
    \includegraphics[width=0.45\textwidth]{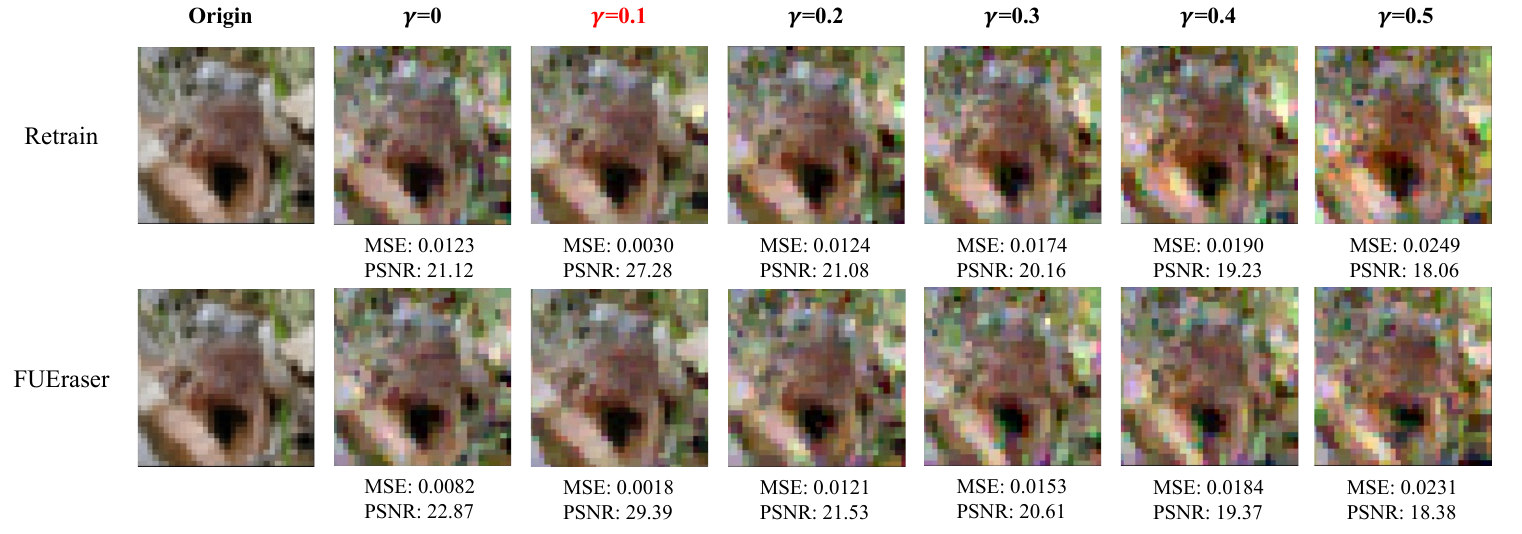} 
    \caption{Effectiveness of FUIA against client unlearning with different values of \(\gamma\).}
    \label{fig_ex_cliun_abl_hyper}
\end{figure}
As previously discussed, \(\nabla^*\) not only represents the gradient information of the forgotten client but also contains noise introduced from other clients during the FL process. Therefore, it is used as an auxiliary term to enhance the gradient features of the target client. To balance the influence of \(\nabla^k\) and \(\nabla^*\), we set \(\gamma = 0.1\) to suppress noise interference and improve reconstruction quality.
To further analyze the effect of \(\gamma\) on FUIA's reconstruction performance, we conduct a set of experiments using the CIFAR-10 dataset. We set the number of data points per client to 1 and randomly choose one client for unlearning. The FUIA is then applied using different values of \(\gamma \in \{0, 0.1, 0.2, 0.3, 0.4, 0.5\}\).
As shown in Figure \ref{fig_ex_cliun_abl_hyper}, when \(\gamma\) exceeds 0.1, the quality of the reconstructed images decreases. The MSE increases, and the PSNR decreases, indicating that excessive use of \(\nabla^*\) introduces detrimental noise. This supports our hypothesis that \(\nabla^*\) should be treated only as auxiliary information. On the other hand, when \(\gamma\) is set too low, the reconstruction metrics also degrade, demonstrating that \(\nabla^*\) does contain useful gradient information related to the forgotten client and should not be ignored. Based on these observations, we adopt \(\gamma = 0.1\) to effectively utilize \(\nabla^*\) as an auxiliary term.

\subsubsection{The Number of data on Each Client}
\begin{figure}[h!]
    \centering
    \includegraphics[width=0.45\textwidth]{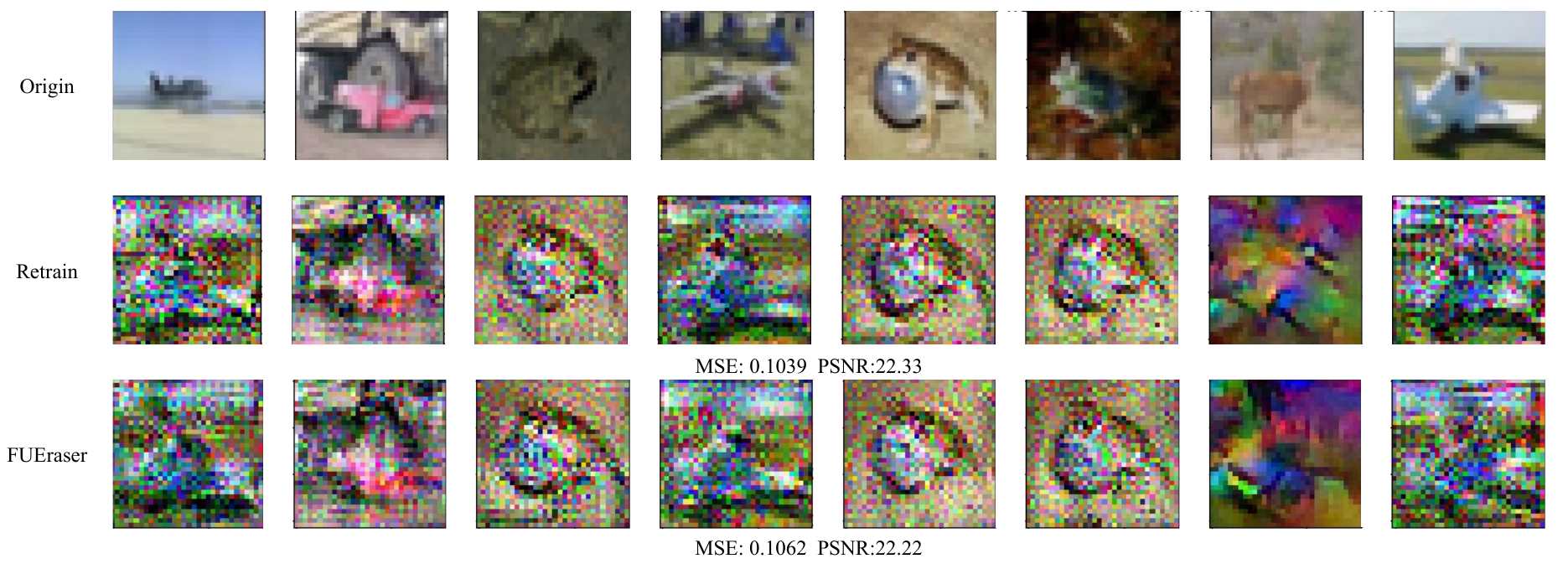} 
    \caption{Effectiveness of FUIA against client unlearning with different numbers of client data.}
    \label{fig_ex_cliun_abl_8un8}
\end{figure}

We further explore the impact of the number of data per client on the effectiveness of the FUIA attack. Specifically, to better present the experimental results, we selected CIFAR-10 as the dataset and set the number of data points per client to 8, then randomly chose one client for unlearning and performed the FUIA attack. The experimental results, shown in Figure \ref{fig_ex_cliun_abl_8un8}, present the outcome of a randomly selected trial. It is evident from the figure that the effectiveness of the FUIA attack significantly decrease. Only a partial reconstruction of the forgotten client data is achieved, and the reconstructed image is relatively blurred. In terms of parameters, the MSE significantly increase compared to the previous results, while the PSNR clearly decrease.
Objectively, it is considerably more challenging to attack client unlearning compared to sample unlearning. Unlearning an entire client causes a larger impact on the model parameters, introducing more significant disturbances. Additionally, accurately collecting gradient information for the entire client is a difficult task, which is also reflected in the previous MUIA experimental results, where MUIA almost failed to reconstruct any meaningful information from the target client. Overall, as the number of forgotten clients increases substantially, the effectiveness of the FUIA is weakened.

\subsection{Ablation Study for FUIA against Class Unlearning}
% \subsubsection{Multi-class Unlearning}
\begin{figure}[h!]
    \centering
    \begin{subfigure}[b]{0.24\textwidth}
        \centering
        \includegraphics[width=\textwidth]{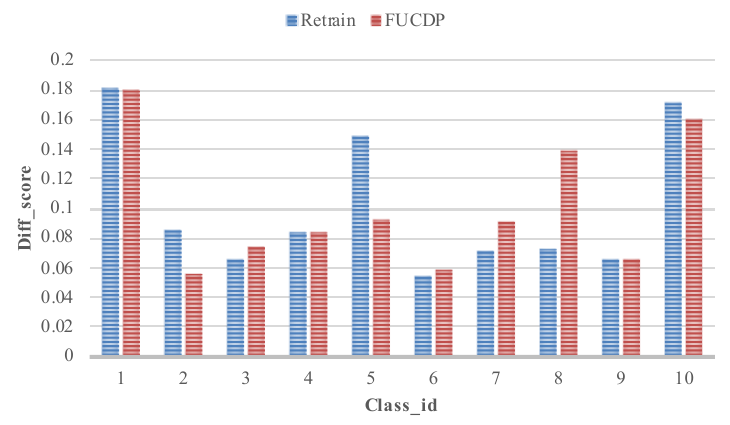} 
        \caption{Unlearning class 0+9}
        \label{fig_ex_claun_abl_class0+9}
    \end{subfigure}
    \hfill
    \begin{subfigure}[b]{0.24\textwidth}
        \centering
        \includegraphics[width=\textwidth]{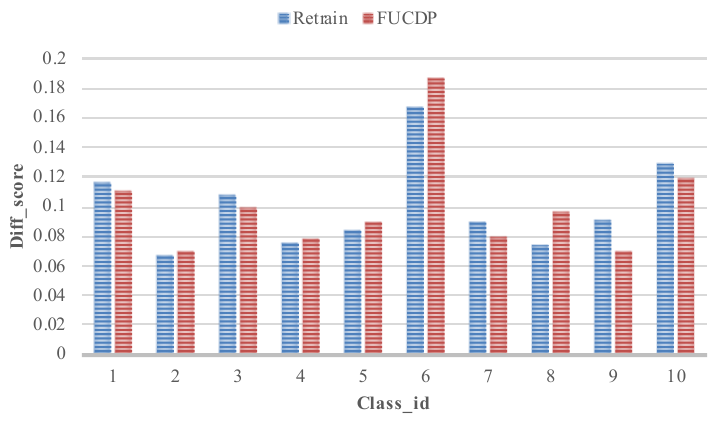} 
        \caption{Unlearning class 0+5+9}
        \label{fig_ex_claun_abl_class0+5+9}
    \end{subfigure}
    \caption{Effectiveness of FUIA against class unlearning with multiple classes unlearning.}
    \label{fig_ex_claun_abl_muticlass}
\end{figure}

To further analyze the impact of unlearning multiple classes on the effectiveness of FUIA attacks, we investigate scenarios where multiple classes are forgotten in a single unlearning process. Specifically, we randomly select 2 and 3 classes for unlearning and then perform attacks using FUIA. The experimental results are presented in Figure \ref{fig_ex_claun_abl_muticlass}, showing two trials. In Figure \ref{fig_ex_claun_abl_class0+9},  when classes 0 and 9 are selected for unlearning, the discrimination scores computed by FUIA indicate that these two classes have the highest scores, successfully identifying the forgotten class labels. Similarly, in Figure \ref{fig_ex_claun_abl_class0+5+9}, when unlearning 3 classes, FUIA is still able to accurately infer the forgotten class labels.  
Furthermore, we conducted extensive experiments to evaluate the attack performance of FUIA in multi-class unlearning scenarios. As shown in Table \ref{table_claun}, when 4 classes are unlearned, there is a 20$\%$ chance that FUIA can not fully infer the labeling of the forgotten category. This may be due to the correlation between the forgotten and remaining classes, leading to a small probability that some remaining classes have slightly higher scores than the forgotten ones. Overall, FUIA remains effective in attacking multi-class unlearning, but its accuracy decreases as the number of forgotten classes increases.

\subsection{Aggregation Method}
\begin{figure}[h!]
    \centering
    \includegraphics[width=0.35\textwidth]{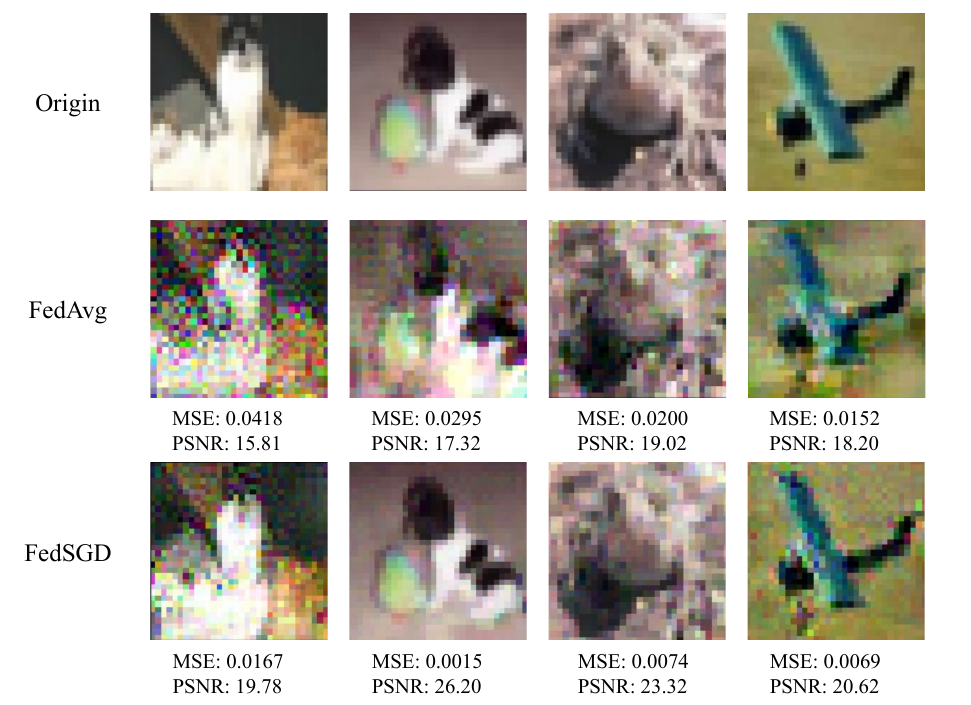} 
    \caption{Effectiveness of FUIA against sample unlearning with different aggregation methods.}
    \label{fig_ex_aggregation}
\end{figure}
We conduct a comparative analysis of the attack performance of FUIA  when using FedAvg and FedSGD as aggregation methods. Specifically, we apply FUIA to attack sample unlearning experiments using the Retraining method aggregated by FedAvg and FedSGD. In the experiments, the attacker obtains local model parameter information via FedAvg, while it can directly access local model training gradient information via FedSGD. The experimental results, shown in Figure \ref{fig_ex_aggregation}, clearly indicate that FUIA achieves significantly better attack performance when FedSGD is used as the aggregation method compared to FedAvg. Moreover, the MSE and PSNR values further confirm this observation.
This phenomenon aligns with our previous analysis. Since FUIA relies on gradient information for attacks, FedSGD enables the attacker to directly and accurately obtain the target gradients, reducing estimation errors and achieving better attack performance compared to FedAvg.

\section{Possible defenses}
\label{defense}
In this section, we draw inspiration from related research \cite{abadi2016deep, zhu2019deep} to design two general defense strategies. In our experiments, we use the CIFAR-10 dataset and choose sample unlearning to demonstrate the defense effectiveness, as FUIA against sample and client unlearning both rely on GIA.

\begin{table}[]
    \caption{Accuracy of models with different defense parameters.}
    \label{table_defense_acc}
    \centering
    \setlength{\tabcolsep}{1.5mm}
    \renewcommand{\arraystretch}{1.5}
    \begin{tabular}{cccc}
    \toprule[1.5pt] % 设置顶线粗细
    Defense                                & Parameter            & Parameter value & Averaged accuracy \\ \hline
    None                                   & -                    & -               & 84.10\%           \\ \hline
    \multirow{4}{*}{Gradient Pruning}      & \multirow{4}{*}{$p$}   & 0.5             & 78.32\%           \\ \cline{3-4} 
                                           &                      & 0.6             & 72.27\%           \\ \cline{3-4} 
                                           &                      & 0.7             & 67.78\%           \\ \cline{3-4} 
                                           &                      & 0.8             & 62.21\%           \\ \hline
    \multirow{4}{*}{Gradient Perturbation} & \multirow{4}{*}{$std$} & 0.001           & 80.64\%           \\ \cline{3-4} 
                                           &                      & 0.003           & 76.43\%           \\ \cline{3-4} 
                                           &                      & 0.005           & 72.33\%           \\ \cline{3-4} 
                                           &                      & 0.009           & 65.32\%           \\
    \bottomrule[1.5pt] % 设置底线粗细
    \end{tabular}
\end{table}

\subsection{Gradient Pruning}
\begin{figure}[h!]
    \centering
    \includegraphics[width=0.45\textwidth]{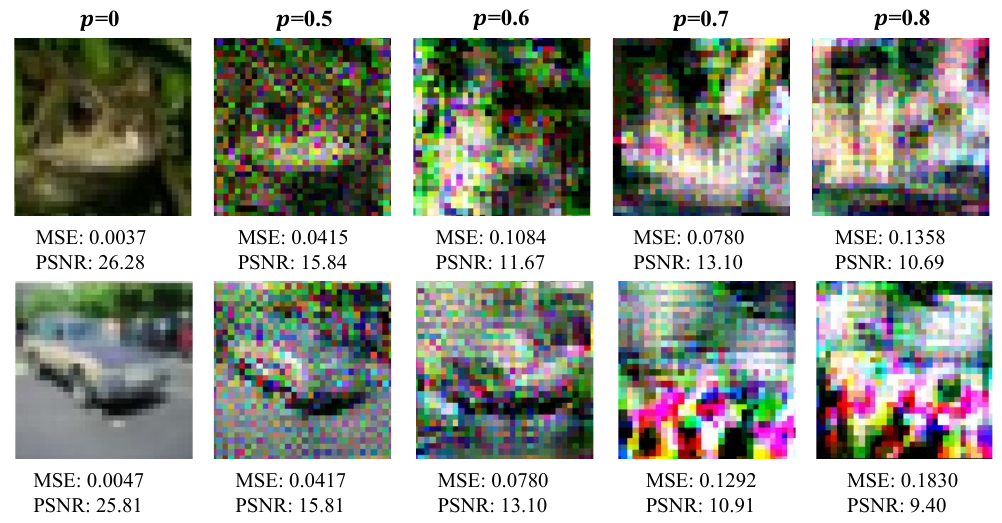} 
    \caption{Defense effects at different pruning ratios.}
    \label{fig_ex_def_pruning}
\end{figure}
During local training on each client, we prune the least significant gradients to obscure gradient information, thereby reducing the effectiveness of FUIA. Specifically, we prune a proportion $p$ of the gradients with the smallest absolute values in each training round. As shown in Figure \ref{fig_ex_def_pruning}, when $p=\{0.5,0.6,0.7,0.8\}$, the reconstructed images by FUIA become increasingly blurry as $p$ increases, with MSE values rising and PSNR values dropping. When $p>0.6$, FUIA fails to reconstruct any critical information about the forgotten data, demonstrating the effectiveness of this method. However, as $p$ increases, the validation accuracy on the test set decreases, as shown in Table \ref{table_defense_acc}, with accuracies of 78.32$\%$, 72.27$\%$, 67.78$\%$, and 62.21$\%$, respectively. To balance defense effectiveness and model performance, we select $p=0.6$ as the pruning ratio, which effectively defends against FUIA while controlling the decline in model performance.

\subsection{Gradient Perturbation}
\begin{figure}[h!]
    \centering
    \includegraphics[width=0.45\textwidth]{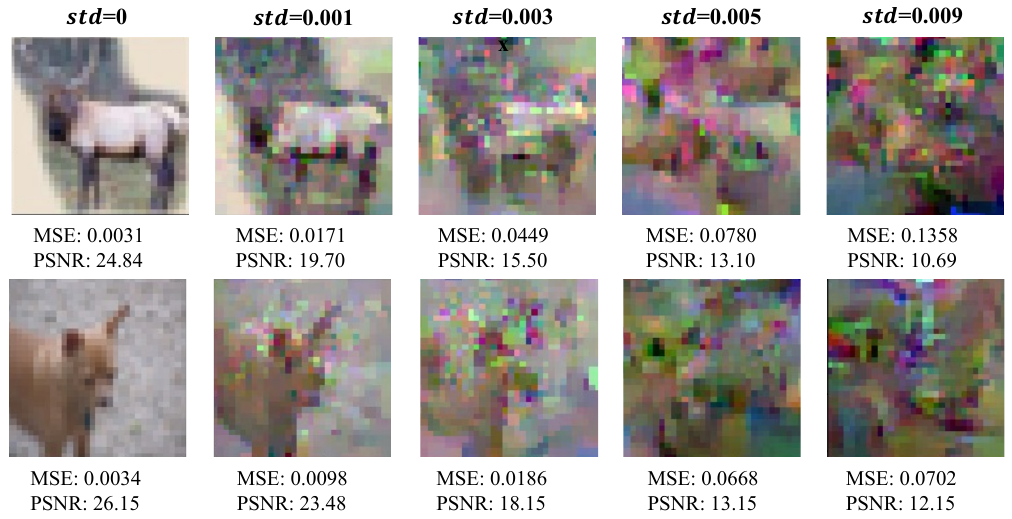} 
    \caption{Defense effects under different noises.}
    \label{fig_ex_def_noise}
\end{figure}
Inspired by differential privacy, we add Gaussian noise to gradients during client training to obscure the target gradients, reducing the effectiveness of FUIA. Specifically, we add Gaussian noise with a mean of 0 and standard deviations ($std$) of $\{0.001, 0.003, 0.005, 0.009\}$. As shown in Figure \ref{fig_ex_def_noise}, as the noise intensity ($std$ value) increases, the reconstructed images by FUIA gradually lose their target features and become increasingly blurred, with MSE values rising and PSNR values dropping. When the $std$ value exceeds 0.003, the reconstructed images by FUIA lose most of the critical features of the forgotten data. However, adding noise also affects model performance, as shown in Table \ref{table_defense_acc}, where validation accuracies under different noise intensities are 80.64$\%$, 76.43$\%$, 72.33$\%$, and 65.32$\%$, respectively. Therefore, we choose a Gaussian noise with $std=0.003$ as the defense parameter, which effectively defends against FUIA while preserving as much model performance as possible.

\section{Conclusion}
\label{conclusion}
In this study, we firstly focus on the privacy leakage issue inherent in FU and propose FUIA.  
FUIA covers the main current FU scenarios and adapts its attack strategy based on different unlearning targets. The core mechanism involves tracking parameter changes during the unlearning process to reconstruct forgotten data. This attack directly contradicts the fundamental goal of FU, which is to completely remove the influence of specific data, thereby exposing inherent privacy vulnerabilities in FU methods.  
To evaluate the effectiveness of FUIA, we conduct extensive experiments demonstrating its ability to infer private information from forgotten data. Additionally, we perform a systematic ablation study to analyze the impact of key variables on attack performance. Furthermore, we explore two potential defense mechanisms to mitigate the security risks posed by FUIA.  
Our work highlight the privacy risks in FU methods and aim to contribute to the development of more secure and robust federated unlearning mechanisms.

% here is the acknowledgement
% \section*{Acknowledgment}
% The authors would like to thank a.

% Don't be bothered with this part =_=!!
\ifCLASSOPTIONcaptionsoff
  \newpage
\fi

% Here goes your Bibtex References!! Uncomment them before using!
\bibliographystyle{IEEEtran}         
\bibliography{fuia_reference}

% Here ends the Paper

\end{document}